\documentclass[a4paper,11pt]{article}
\usepackage{amssymb, amsmath}

\newtheorem{theorem}{Theorem}[section]

\newtheorem{lemma}[theorem]{Lemma}
\newtheorem{proposition}[theorem]{Proposition}

\numberwithin{equation}{section}

\title{On thermodynamic states of the Ising model on scale-free graphs}
\author{Yuri Kozitsky\\[.5cm]{Institute of Mathematics,}\\[.2cm] Maria Curie-Sklodowska
University,\\[.2cm] 20-031 Lublin, Poland\\[.4cm] {jkozi@hektor.umcs.lublin.pl}}

\begin{document}

\maketitle

\begin{abstract}

There is proposed a model of  scale-free random graphs which are locally close to the uncorrelated
complex random networks with divergent $\langle k^2 \rangle$ studied in
e.g. S. N. Dorogovtsev {\it et al}, Rev. Mod. Phys. {\bf 80}, 1275 (2008).
It is shown that the Ising model on these graphs with interaction intensities of arbitrary signs with probability one is in a paramagnetic state at sufficiently high finite values of the temperature.
For the same graphs, the bond percolation model with probability one is in a nonpercolative state for positive values of  the percolation probability.
Possible extensions are discussed.

\end{abstract}

\section{Introduction}
\label{1S}

In statistical physics, a paradigm model
 for cooperative phenomena is the Ising model of interacting spins attached to
the vertices of a graph ${\sf G} = ({\sf V}, {\sf E})$. This model is described by the Hamiltonian
 \begin{equation}
   \label{1}
  H = -  \sum_{\{x, y\}\in {\sf E}}J_{xy} \sigma_x \sigma_y - h \sum_{x\in {\sf V}}  \sigma_x,
 \end{equation}
where $J_{xy}$ are interaction intensities, $h$ is an external
field, and the spins $\sigma_x$ take values $\pm 1$.   The sums run over the sets of edges
${\sf E}$ and vertices ${\sf V}$ of the graph ${\sf G}$,  which is supposed to be connected and countably infinite.
In view of the latter fact, the Hamiltonian (\ref{1}) has no direct mathematical meaning and is used as a formula for local Hamiltonians, $H_{\sf \Lambda}$, which one
obtains by restricting the sums to finite subsets ${\sf \Lambda} \subset {\sf V}$ and ${\sf E}_{\sf \Lambda} \subset {\sf E}$, where
${\sf E}_{\sf \Lambda}$ is the collection of edges with both ends in ${\sf \Lambda}$. Then the thermodynamic properties of (\ref{1})
are studied in the limit ${\sf \Lambda} \to {\sf V}$. In a more sophisticated theory \cite{Ge}, a thermodynamic state
is a Gibbs measure, and a phase transition is associated with the possibility for multiple thermodynamic states to exist at the same values of the temperature and $h$.

Obviously, the properties of the model (\ref{1}) are closely related to those of the underlying graph.
The case of a special interest is where this graph is random, with
vertices having arbitrary
number of neighbors. Then the graph itself can be characterized by the properties of the corresponding  Ising model.
Perhaps in view of this fact,
methods of statistical physics are widely used also in the study of the underlying random graphs. A long list of publications on this topic can be found in
the review articles \cite{AlB,Dor,N}.

For a vertex $x$, the degree $k(x)$ is defined as the number of neighbors of $x$, i.e. the number of $y\in {\sf V}$ such that
$\{x,y\} \in {\sf E}$.
If ${\sf G}$ is random, the degree of each $x\in {\sf V}$ is an integer valued random variable and hence is characterized by the
probability distribution
\begin{equation}
  \label{2}
 {\rm Prob} \{k(x) = k\} = p_k (x),  \qquad k\geq 1, \quad x \in {\sf V}.
\end{equation}
In simple cases, the degrees $k(x)$ are independent and identically distributed, which in particular means
that the probabilities $p_k(x)$ are the same for all $x$. If $p_k = C k^{-\lambda}$, at least for big enough $k$ and an appropriate
$C>0$, the corresponding graph is said
to be {scale-free}, c.f. \cite{Bar}. In this case, the $m$-th moment
\[
\langle k^m \rangle = \sum_{k\geq 1} k^m p_k , \qquad m\geq 1,
\]
exists provided $\lambda > m+1$. Notably, $\langle k^2 \rangle = + \infty$ for many real-world
networks \cite{N}.

An example of  a random graph with independent $k(x)$ is a Galton-Watson tree, which is the genealogic tree
of a Galton-Watson branching process \cite{As,Yuval}. In more complex models,
the graph ${\sf G}$ is obtained, more or less explicitly, as the limit  of an
increasing sequence of finite random graphs, $({\sf V}_n, {\sf E}_n)$, $n\in \mathbb{N}$, see, e.g. \cite{BCK,BK,Dembo,Dommers}.
In the physical literature, such constructions  are rather informal that leaves certain possibilities for
different interpretations, which we are going to use in this work. In 2002, almost simultaneously there appeared two papers \cite{D,L}
where the model (\ref{1}) on a scale-free graph was studied.
The starting point is a scale-free distribution $\{p_k\}_{k\in \mathbb{N}}$, which ``characterizes the entire net", whereas the ``local structure
of connections" is characterized by ``the distribution of the number of connections of the nearest neighbor of a vertex", which is taken to be $k p_k/\langle k \rangle$, see the beginning of the first section in \cite{D}.
In \cite{L}, a similar assumption was made, see equation (3) thereof, and also  (1) in \cite{Dor}. In the latter article, such graphs are called
{\it complex networks} in contrast to the classical models called
{\it simple networks,} one of which is presumably the Galton-Watson tree mentioned above. In both papers \cite{D,L},
it was found that the model (\ref{1}) with $h=0$, ferromagnetic interactions $J_{xy} = J >0$, and $\langle k^2 \rangle = + \infty$
is in an ordered (ferromagnetic) state at all
temperatures\footnote{In contrast to
regular graphs, e.g. crystal lattices, the Ising model on sufficiently `dense' non-random graphs  can be in
an ordered state at all temperatures, see the example in \cite{KK}.} $T>0$.
Recently, in \cite{Dembo,Dommers} the mathematical construction of a random graph model was presented,
in which the mentioned properties were realized.
For this model, the conclusion of \cite{D,L}  can be mathematically proven, see \cite{Rly} and Subsection \ref{conc} below. The aim of this work is
to construct ``uncorrelated  complex networks" with $\langle k^2 \rangle = \infty$,
consistent with the assumptions made in \cite{Dor,D,L},
for which the model (\ref{1}) with interaction intensities of arbitrary signs can be in a paramagnetic state at finite temperatures.
In Section \ref{2S}, we perform this construction
and obtain a graph model, which is a combination of the one used in \cite{Dembo,Dommers} and of the ordinary Galton-Watson tree.
In Section \ref{3S}, we prove Theorem \ref{Tm} the main result of which is that the model (\ref{1}) on the introduced graphs,  with any signs of $J_{xy}$,
can be in a paramagnetic state at finite temperatures and $h=0$, provided $\langle k^{1+ \alpha} \rangle <\infty$ for some $\alpha >0$.
This covers the case of scale-free distributions with $\lambda \in (2 , 3)$. The proof of Theorem \ref{Tm} relies upon two lemmas, proven in Section
\ref{4S}. The result of Theorem \ref{Tm} and the graph model introduced in Section \ref{2S} are discussed in the concluding Section \ref{5S}, where we give also an extension of the results of \cite{D,L} concerning the model (\ref{1}) with interaction intensities of arbitrary signs. Except possibly for Section \ref{4S}, the paper is reasonably self-contained and accessible to non-mathematicians.

\section{The graph models}
\label{2S}

Since the Galton-Watson tree will serve us as a prototype, we beging by presenting this model.
More on this topic can be found in \cite{As,Yuval,LyP}.

\subsection{The Galton-Watson tree}

To exclude the possibility of obtaining a finite tree, or a tree of bounded degree, we shall suppose that the basic  random variable
$X$  takes positive integer values $k\in
\mathbb{N}$ with probability ${\rm Prob}(X= k) = p_{k+1}$
such that
\begin{eqnarray}
  \label{PQ}
  p_k \in [0,1),  \quad  & & \sum_{k\geq 2}p_k =1, \qquad \  \sum_{k= 2}^n p_k <1, \ \ {\rm for} \ {\rm any}
 \ n\in \mathbb{N},   \\[.2cm]
 & &    a:= \sum_{k\geq 2} k p_k  < \infty. \nonumber
\end{eqnarray}
One observes that
\begin{equation}
 \label{Aa}
a > 2,
\end{equation}
which readily follows form the properties of $p$ assumed in (\ref{PQ}).
Set $L_0 = 1$, $L_1=X$, and
\begin{equation}
  \label{L}
L_{n} = \sum_{j=1}^{L_{n-1}} X_{n,j},
\end{equation}
where all $X_{n,j}$ are independent copies of $X$, see Chapters II and III in
\cite{As} or  Chapter 5
in \cite{Yuval}. Then $L_n$ is the number of
individuals in generation $n$ of a Galton-Watson branching process.
The process starts with one individual, and each
individual in generation $n$ produces offsprings at random,
independently and with the same distribution
as $X$. The individuals existing in all generations form the vertex
set ${\sf V}$ of the corresponding (genealogic) Galton-Watson graph ${\sf G}$,
with undirected edges $\{x,y\}\in{\sf E}$ connecting each individual
and its
offsprings.
Under the assumptions made in  (\ref{PQ}), the corresponding process does not extinct and $L_n \rightarrow + \infty$
with probability one. Hence,
${\sf G}$ is an infinite tree -- an acyclic connected graph,
which  has no leaves, except possibly for
the root $o$, the  degree of which
coincides with the number of the corresponding offsprings $L_0$. The
degree of any other vertex equals the number of its offsprings plus
one. By construction, the vertex
degrees are independent identically distributed random variables such that
\begin{equation}
  \label{3b}
{\rm Prob}\{ k(o) = k\} = p_{k+1}, \qquad {\rm Prob}\{ k(x) = k \}
= p_k, \quad x\in {\sf V}\setminus \{o\}.
\end{equation}
Now let us fix the graph-theoretical and  probabilistic terminology used in this article.
For a graph ${\sf G}=({\sf V}, {\sf E})$ and vertices $x,y\in {\sf V}$, a {\it path} $\vartheta (x,y)$ is a sequence
$\{x_0, x_1, \dots, x_n\}$, $n\in \mathbb{N}$, such that: (a) $x_0 =x$, $x_n = y$; (b) $\{x_{l}, x_{l+1}\}\in {\sf E}$; (c)
neither of $x_0, \dots ,x_{n-1}$ can be repeated. The length of the path is $\|\vartheta(x,y)\| =n$; $x$ is its origin and $y$ is terminus. A {\it ray} with origin $x_0$ is an infinite sequence of distinct vertices
$\{x_0, x_1, \dots, x_n, \dots\}$ with property (b).
A
graph ensemble $\mathcal{G}$ is the collection of all possible realizations, e.g., as
just described. A random graph model is a triple $(\mathcal{G}, \mathcal{F}, P)$, where
$\mathcal{F}$ is a certain family of subsets of $\mathcal{G}$ and $P: \mathcal{F}\to [0,1]$
is probability. For the Galton-Watson tree, $P$ is defined by the collection $p=\{p_k\}$.
For $\mathcal{A}\in \mathcal{F}$,
$P(\mathcal{A})$ is the probability of the event:  ``{the graph chosen at random from $\mathcal{G}$ belongs to $\mathcal{A}$}''.
Some events
can occur with probability either zero or one  -- the celebrated zero-one law.
In the Galton-Watson tree, if $\mathcal{A}$ is the set of all realizations with $k(o) =1$, then $P(\mathcal{A})=p_1$.
We say that a property of a random graph holds with {\it probability one} if this property
is possessed by  all of the elements of some $\mathcal{A}$ such that
$P(\mathcal{A})=1$.
Note that this $\mathcal{A}$ need not be the whole ensemble $\mathcal{G}$. Usually, when one deals with models with independent vertex degrees,
 the triple $(\mathcal{G}, \mathcal{F}, P)$ does not
appear explicitly. This will also be the case in the rest of this paper.

\subsection{The configuration  model}

Let $p:= \{p_{k}\}_{k\geq 2}$ be as in (\ref{PQ}), and
\begin{equation}
 \label{Aa1}
\hat{p}_k := kp_k /a, \qquad k\geq 2.
\end{equation}
``Uncorrelated complex networks`` were `defined' in \cite{Dor,D,L} by means of the following conditions.
The main one is that ``the neighbor of a vertex has degree $k$ with probability $\hat{p}_k$", which can be interpreted as the
condition that
 in every path $\{x_0, x_1, \dots , x_n\}$, the probability that $k(x_l)$
takes value $k$ is $\hat{p}_k$. For convenience, we shall call it a {\it path property}.
Another important assumptions are: (a) all $k(x)$
are ``uncorrelated" (in fact, independent); (b) two paths $\{x,x_1, \dots, y\}$ and
$\{x, x_1', \dots, z\}$, with $x_1 \neq x_1'$ and $y\neq z$, never intersect each other, i.e., the graph is a
tree\footnote{In \cite{Dor,D}, the graphs were called ``tree-like" if the intersections are negligible.  Mathematical definitions
of tree-like graphs can be found in \cite{Dembo,Dommers,Rly}.}.
These assumptions are consistent with the following mathematical models.
The first one is the ordinary Galton-Watson tree with $p$ replaced by $\hat{p}$, in \cite{Dembo} it is
${\sf T}(\rho,\infty)$, see page 567 threof.
A slightly different version is the so called {\it configuration model}, c.f. page 1279 in \cite{Dor}.
This model is a rooted tree with independent vertex degrees such that
\begin{equation}
  \label{RT}
{\rm Prob} (k(o) =k) = p_k, \qquad  {\rm Prob} (k(x) =k) = \hat{p}_k.
\end{equation}
This model\footnote{Also models which
{\it locally converge} to ${\sf T}(P, \rho, \infty)$, cf. Definition 2.1 in \cite{Dembo}.}  was studied in \cite{Dembo,Dommers}, in
\cite{Dembo} it appears as ${\sf T}(P, \rho, \infty)$. The main characteristic feature of both ${\sf T}(P, \rho, \infty)$ and ${\sf T}(\rho, \infty)$ is that {\it each} path
$\vartheta (x,y)$, $x\neq o$, has the path property
described above. The inconsistency with the root degrees does not change the
global behavior of the graph, and hence of the corresponding model (\ref{1}).

The construction of the configuration model can  be visualized as follows.
Consider a countable set of `points' (a configuration),
each of which is given a random mark\footnote{According to \cite{Dor}, a mark is the number of stubs.}
$k\in \mathbb{N}$ with probability distribution $p$ as in (\ref{PQ}).
The marks of different points are independent. Then the graph is obtained as follows.
The root degree is drawn from $\mathbb{N}$ according to ${\rm Prob} (k(o)=k) = p_k$.
If the result is $m$, the root receives $m$ neighbors -- `particles' independently drawn
from the configuration with probability proportional to the particle mark, i.e., $C k$.
Then the (unconditional) probability of drawing some particle, which bears mark $k$, is $C kp_k$ with $C=1/a$
since the total probability should sum up to one. In the result,
each root neighbor has degree $k$ with probability $\hat{p}_k$.
In the same way one picks the next-neighbors of $o$ keeping in mind that the only common element of the neighborhoods
of different vertices is the root itself.
This procedure is continued ad infinitum until the whole tree is constructed.
An important observation here is that, except for the root degree,
the distribution of $k(x)$ is given by $\hat{p}$, whereas $p$ itself is the distribution of marks -- auxiliary objects
used in the graph construction.

\subsection{The proposed model}
\label{grf}
Here the main object of our study in this article is introduced. This is a random tree model, also consistent with the assumptions of \cite{Dor,D,L}.
As in the case of the Galton-Watson tree, we beging with the description of the corresponding branching process.  Let
$p$, $\hat{p}$, and $a$ be as in (\ref{PQ}) and (\ref{Aa1}). An additional parameter is $s\in \mathbb{N}$.
The whole population falls into
two types -- `distinguished' and `ordinary' individuals. At the beginning, ordinary individuals are absent and there is one distinguished individual, which independently produces
$m$ offsprings with probability $\hat{p}_m$. If $m\leq s$, all of them are set to be distinguished. For $m>s$, $s$ offsprings are set to be distinguished and the remaining $m-s$ ones are set to be  ordinary. Each ordinary individual produces independently $k$ ordinary individuals with probability $p_{k+1}$. Each distinguished individual, other than the initial one, produces independently $m$ offsprings with probability $\hat{p}_{m+1}$. As above, if
$m\leq s$, then all of the offsprings are distinguished. For $m>s$, $s$ offsprings are distinguished and the remaining $m-s$ ones are ordinary.
This process is repeated ad infinitum. For $n\in \mathbb{N}_0$, by $\widehat{L}_n$ and $L_n$ we denote  the number of distinguished and ordinary
individuals in generation $n$, respectively. Let also $\widehat{X}$ and $X$ be independent and such that ${\rm Prob}(\widehat{X} = m) = \hat{p}_{m+1}$ and ${\rm Prob}({X} = m) = {p}_{m+1}$, $m\in \mathbb{N}$. Then we have, cf. (\ref{L}),
\begin{eqnarray}
  \label{Lh}
\widehat{L}_n & = & \sum_{j=1}^{\widehat{L}_{n-1}}\left[  \widehat{X}_{n,j} -  \left( \widehat{X}_{n,j} -s\right)_+\right], \\[.2cm] L_n & = & \sum_{j=1}^{L_{n-1}} X_{n,j} + \sum_{j=1}^{\widehat{L}_{n-1}} \left( \widehat{X}_{n,j} -s\right)_+, \nonumber
\end{eqnarray}
where, for an integer $\varkappa$, we write $\varkappa_+ = \max\{0, \varkappa\}$, and $\widehat{X}_{n,j}$ and $X_{n,j}$ are independent copies of
$\widehat{X}$ and $X$, respectively. From the point of view of ordinary particles, those represented
by the second term in the expression for $L_n$ are {\it immigrants}, and their number in generation $n$ is
\begin{equation}
  \label{Lh1}
  Y_n = \sum_{j=1}^{\widehat{L}_{n-1}} \left( \widehat{X}_{n,j} -s\right)_+.
\end{equation}
Then the total number of individuals in generation $n$ can be written in the form
\begin{equation}
  \label{Lh2}
  \widetilde{L}_n := \widehat{L}_{n} + L_n = \sum_{j=1}^{\widehat{L}_{n-1}} \widehat{X}_{n,j} + \sum_{j=1}^{L_{n-1}} X_{n,j}.
\end{equation}
Since each distinguished individual can have at most $s$ distinguished offsprings, we have that
\begin{equation}
  \label{Lh3}
  \widehat{L}_n \leq s^n.
\end{equation}
The  graph in question is
the genealogic tree of the process just described.  Its construction can be visualized as follows.
The graph is a rooted tree with ${\rm Prob} (k(o)=k) = \hat{p}_{k+1}$. Let $k(o) =m\leq s$. Then we draw $m$ neighbors of $o$ as in the configuration model. Denote them
$x_1, \dots ,x_m$; by construction ${\rm Prob} (k(x_l)=k) = \hat{p}_k$ for all $l=1, \dots m$.
If $m>s$, we draw $s$ neighbors $x_1, \dots ,x_s$ of $o$ as in the configuration model. The remaining neighbors $y_{s+1}, \dots , y_m$ are set to be the roots of independent Galton-Watson trees with probability distribution $p$, which in particular means that ${\rm Prob}(k(y_l) = k)= p_k$, $l=s+1, \dots , m$. This procedure is continued ad infinitum.
For $s>1$, the obtained tree contains a subtree, comprised by  an
 infinite number of {\it distinguished} rays $\{o, x_1, x_2, \dots , x_n, \dots \}$ with
the mentioned path  property ``the neighbor of $x_{n-1}$ has degree  $k$  with probability $\hat{p}_k$".
The set of vertices of this subtree is contained in that of a Cayley tree with root $o$ and branching number $s$.
For $s=1$, our graph is a size-biased Galton-Watson tree, see \cite[pp. 407--412]{Yuval} or \cite{LyP},
in which there is only one distinguished ray.

In the remaining part of the article  we use the following nomenclature. By ${\bf GW}(s,p)$, $s\in \mathbb{N}_0$,
we denote the model constructed in this subsection. Then ${\bf GW}(1,p)$ and ${\bf GW}(0,p)$ are is the size-biased and the ordinary Galton-Watson trees, respectively.

\section{The thermodynamic states of the Ising model}

\label{3S}

\subsection{The thermodynamic states}
\label{SS1}

Let ${\sf G}= ({\sf V}, {\sf E})$ be a general tree with root $o$.
Given $x\in {\sf V}$, the distance $\rho(o,x)$ is the length of the path $\vartheta(o,x)$. For
$n\in \mathbb{N}_0$, by ${\sf S}_n$ we denote the collection of vertices $x$ such that $\rho(o,x) =n$.
Put
\[
{\sf V}_n = \bigcup_{m=0}^n {\sf S}_m,
\]
and
let ${\sf E}_n$  be the set
of all edges with both ends in ${\sf V}_n$. Let also  ${\sf
E}^{ b}_n$ be the set of  $\{x,y\}$ such that $x\in {\sf S}_n$ and
$y\in {\sf S}_{n+1}$.

By  $\sigma_n$ we denote the configuration of spins in ${\sf V}_n$, that is, $\sigma_n= \{ \sigma_x: x \in {\sf V}_n \}$.
In the sequel, we suppose that
\begin{equation}
  \label{Aq}
  J:= \sup_{\{x,y\}\in {\sf E}} |J_{xy}| < \infty.
\end{equation}
The Gibbs probability distribution of
configurations $\sigma_n$  at temperature $T$ and $h=0$ is the following probability measure
\begin{equation}
  \label{5}
\pi_n (\sigma_n|\xi) = \frac{1}{Z_n (\xi)} \exp\left(
\sum_{\{x,y\}\in{\sf E}_n} (K_{xy}\sigma_x \sigma_y +K) +
\sum_{\{x,y\}\in{\sf E}^{ b}_n} K_{xy} \sigma_x \xi_y  \right),
\end{equation}
where
\begin{equation}
  \label{Aq1}
  K_{xy} := \beta J_{xy}, \qquad K:= \beta J, \qquad \beta := 1/k_B T.
\end{equation}
Note that $K_{xy}$ can have arbitrary signs, whereas  $K$ is positive. In (\ref{5}),
the first (resp. second) summand in $\exp(\cdots )$
corresponds to the interaction of the spins in ${\sf V}_n$ with each
other (resp. with the spins $\xi_y$ fixed outside ${\sf V}_n$). One observes
that the latter interaction  involves only $\xi_y$ with $y\in
{\sf S}_{n+1}$, which constitute the outer boundary of ${\sf V}_n$. For technical reasons, we add positive constants to the spin-spin interaction
along each edge in ${\sf E}_n$, c.f. (\ref{1}).
The partition function is then
\begin{equation}
  \label{Z}
Z_n (\xi) = \sum_{\sigma_n} \exp\left(  \sum_{\{x,y\}\in{\sf
E}_n} (K_{xy}\sigma_x \sigma_y +K) +   \sum_{\{x,y\}\in{\sf E}^{ b}_n} K_{xy} \sigma_x
\xi_y  \right),
\end{equation}
where the summation is taken over all $\sigma_x= \pm 1$, $x\in{\sf V}_n$.

Given $z\in {\sf V}$, let $n_z\in \mathbb{N}_0$ be such that $z\in {\sf S}_{n_z}$, i.e., $n_z = \rho(o,z)$. For these $z$, $n_z$, and for $n> n_z$,
\begin{equation} \label{21}
M_{n,z} (\xi) = \sum_{\sigma_n} \sigma_z \pi_n (\sigma_n|\xi) = \sum_{\sigma_z =\pm 1} \sigma_z \varrho_{n,z} (\sigma_z |\xi) = \varrho_{n,z} (1 |\xi)- \varrho_{n,z} (- 1|\xi)
\end{equation}
is the magnetization at $z\in {\sf V}_n$ in the state $\pi_n(\cdot|\xi)$
(\ref{5}). Here, for $\sigma_z=\alpha$,
\begin{equation}
  \label{Aq2}
\varrho_{n,z}(\alpha |\xi) = \sum_{\sigma_n: \ \sigma_z = \alpha}\pi_n (\sigma_n|\xi), \qquad \alpha = \pm 1,
\end{equation}
where the summation is taken over all $\sigma_n$ with the fixed $\sigma_z=\alpha$.
According to the theory of Gibbs states \cite{Ge}, the global thermodynamic state is unique if, for all $z$, the limits
\begin{equation}
  \label{Aq3}
  \varrho_z (\alpha) := \lim_{n\to +\infty} \varrho_{n,z}(\alpha |\xi)
\end{equation}
exist and are independent of $\xi$. In this case, by (\ref{21}) the global magnetization $M_z$ exists and
\[
M_z = \lim_{n\to +\infty} M_{n,z} (\xi) = \varrho_z (1) - \varrho_z (-1).
\]
As $h=0$, we have
\begin{equation}
  \label{Aq4}
  M_{n,z} (\xi) = - M_{n,z} (-\xi),
\end{equation}
and hence $M_z = - M_z$. Therefore, the uniqueness of thermodynamic states occurs if and only if, for all $z\in {\sf V}$ and $\xi$,
\begin{equation}
  \label{7}
M_z =  \lim_{n\to +\infty} M_{n,z} (\xi) =0.
\end{equation}
In this case, we say that the model is in a {\it paramagnetic state}.

If the underlying tree ${\sf G}$ is random, then the measure (\ref{5}) is also random, and hence
(\ref{7}) is a random event obeying the zero-one law. If the corresponding probability is one, we say that
the model (\ref{1}) on ${\sf G}$ is in a paramagnetic state  with probability one.

\subsection{The main statement}

Let $p$ and $\hat{p}$ be as in
(\ref{PQ}) and (\ref{Aa1}). For $\alpha \in (0,1)$, we set
\begin{equation}
  \label{Log}
 b_\alpha =  \sum_{k\geq s+1} (k-s)^\alpha \hat{p}_k ,  \qquad   b= \sum_{k\geq 2} k p_k \ln k ,
\end{equation}
and also
\begin{equation}
\label{a}
K_c = \frac{1}{4} \ln \frac{a}{a-1}, \qquad \widehat{K}_c = \frac{1}{4} \ln \frac{s^\gamma +1}{s^\gamma}, \quad \ \gamma = 1/{\alpha}  .
\end{equation}
In the statement below we describe the Ising model (\ref{1}) on ${\bf GW}(s,p)$ graphs with $p$ obeying (\ref{PQ}).
\begin{theorem}
 \label{Tm}
For $s=0$, the Ising model is in a paramagnetic state whenever $K< K_c$. For $s=1$, the same holds under the condition
$b<\infty$. For $s\geq 2$, the Ising model is in a paramagnetic state whenever $b_\alpha < \infty$ for some $\alpha \in (0,1)$ and $K< \min\{K_c; \widehat{K}_c\}$.
\end{theorem}
In Section \ref{5S}, we discuss the above statement in detail.
Its proof is based on two lemmas proven in the next section.
In the first lemma, we describe the model (\ref{1}) on a general tree, for which the corresponding quantities were introduced in Subsection \ref{SS1}.
\begin{lemma}
 \label{Pn}
Let $K>0$ be such that
\begin{equation}
 \label{Qq}
q(K) := \exp(4K) - 1<1.
\end{equation}
Then for each $z\in {\sf V}$, any $n> n_z$, and arbitrary $\xi$ and $\eta$, we  have that
\begin{equation}
\label{R2}
\left\vert M_{n,z} (\xi) - M_{n,z} (\eta) \right\vert \leq 2 [q(K)]^{n-n_z}|{\sf S}_n|.
\end{equation}
\end{lemma}
In the next lemma, we describe the model ${\bf GW}(s,p)$ introduced in Subsection \ref{grf}.
\begin{lemma}
  \label{grflm}
Let $L_n$ be as in (\ref{Lh}), $s\geq 2$, and $b_\alpha < \infty$ for some $\alpha \in (0,1)$. Then, for every $c$ obeying $c> s^\gamma$ and $c\geq a-1$, with probability one
\[
c^{-n} L_n \to W\in [0,\infty), \qquad  {\rm as} \quad n \to \infty.
\]
\end{lemma}
\vskip.1cm \noindent
{\it Proof of Theorem \ref{Tm}:}
For \underline{$s=0$}, by construction $|{\sf S}_n|=L_n$ is the number of
individuals in generation $n$, see (\ref{L}), and
$a-1$ is the mean number of offsprings
in the corresponding Galton-Watson process.
Then the limit
\begin{equation}
  \label{Aq5}
 \lim_{n\to +\infty} \frac{L_n}{(a-1)^n}
\end{equation}
with probability one exists and is finite, see (\ref{Aa}) and Proposition 1.3 in \cite[page 20]{As}.
If
$K< K_c$, then $q(K) < q(K_c)= 1/(a-1) <1$, and hence we can apply (\ref{R2}), which yields that
 with probability one
\begin{equation}
  \label{Aq5a}
\left[q(K) (a-1) \right]^n \frac{L_n}{(a-1)^n} \to 0, \qquad n \to +\infty,
\end{equation}
which in turn by (\ref{R2}) yields (\ref{7}), and hence the proof for this case.

For \underline{$s=1$}, by (\ref{Lh3}) we have $\widehat{L}_n =1$, and hence $|{\sf S}_n| = \widetilde{L}_n = 1 + L_n$.
Since $b<\infty$, with probability one we have
\begin{equation}
  \label{Aq5b}
  \sum_{n\geq 1} (a-1)^{-n} Y_n < \infty,
\end{equation}
see Proposition 6.2 in \cite[page 50]{As}. Then by Theorem 6.1 in \cite[page 50]{As}, ${\widetilde{L}_n}/{(a-1)^n}$ with probability one tends to a finite limit. Then the proof follows as in the case of $s=0$.

For \underline{$s\geq 2$}, the random variables $Y_n$ with different $n$ are no more identically distributed, and hence
$b<\infty$ is not enough to get (\ref{Aq5b}). Instead we use a more restrictive condition $b_\alpha < \infty$, under which we get,
 see Lemma \ref{grflm}, that $c^{-n} |{\sf S}_n| = c^{-n}\widetilde{L}_n = c^{-n} \widehat{L}_n + c^{-n}L_n$ with probability one  tends to a finite limit since $c> s$, cf. (\ref{Lh3}). Note that
\[
K(c) := \frac{1}{4} \ln \frac{c+1}{c} \leq \min\{K_c ; \widehat{K}_c\}.
\]
Hence, for $K < K(c)$, we have $q(K) < 1/c$, and the proof follows as in (\ref{Aq5a}).

\section{The proof of the lemmas}

\label{4S}

\subsection{The proof of Lemma \ref{Pn}}

Let ${\sf G}=({\sf V}, {\sf E})$ be a general tree. For a path $\vartheta$,
by ${\sf E}_\vartheta$ we denote the set of edges $\{x_{l-1}, x_l \}$ of $\vartheta$.
Since ${\sf G}$ is a tree, there exists exactly one path
$\vartheta(x,y)$ for any $x$ and $y$. Recall that ${\sf E}_n$ stands for
the set of edges with both ends in ${\sf V}_n$. Thus, ${\sf G}_n = ({\sf V}_n, {\sf E}_n)$ is a
finite graph; ${\sf G}' = ({\sf V}', {\sf E}')$ is called a subgraph of ${\sf G}_n$ if ${\sf V}' \subset {\sf V}_n$
and ${\sf E}' \subset {\sf E}_n$. A path $\vartheta$ is said to be in ${\sf G}'$ if ${\sf E}_\vartheta \subset {\sf E}'$. Two vertices $z, x\in {\sf V}'$ are said to be {\it disconnected} in ${\sf G}'$
if there is no path $\vartheta (z,x)$ in ${\sf G}'$.

By (\ref{21}) and (\ref{5}), employing replica spins $\tilde{\sigma}$ we obtain the following Meyer-like expansion
\begin{eqnarray}
  \label{M1}
 & & M_{n,z} (\xi) - M_{n ,z} (\eta) = \sum_{\sigma_n, \tilde{\sigma}_n}
 \left(\sigma_z - \tilde{\sigma}_z \right) \pi_n (\sigma_n|\xi)
 \pi_n (\tilde{\sigma}_n|\eta) \\[.2cm] & &
 \qquad = \frac{1}{Z_n (\xi)Z_n (\eta)}\sum_{\sigma_n, \tilde{\sigma}_n} \left(\sigma_z - \tilde{\sigma}_z \right)
 \prod_{\{x,y\}\in {\sf E}_n} ( 1 + \Gamma_{xy}) \Psi_n(\xi, \eta) \nonumber \\[.2cm]
 & & \qquad = \frac{1}{Z_n (\xi)Z_n (\eta)}\sum_{{\sf E}' \subset {\sf E}_n} \sum_{\sigma_n, \tilde{\sigma}_n}
\left(\sigma_z - \tilde{\sigma}_z \right)
\Gamma ({\sf E}') \Psi_n(\xi, \eta), \nonumber
\end{eqnarray}
where
\begin{eqnarray}
  \label{M2}
 & & \qquad \qquad \quad \Gamma({\sf E}')  =   \prod_{\{x,y\}\in {\sf E}'}  \Gamma_{xy},  \nonumber \\[.2cm]
& &  \Gamma_{xy}
  =   \exp\left[ \left(K_{xy}\sigma_x \sigma_y +K \right) + \left(K_{xy}\tilde{\sigma}_x \tilde{\sigma}_y +K \right)\right] -1 , \qquad \\[.2cm]
& & \qquad \Psi_n(\xi, \eta)   =   \prod_{\{x,y\}\in{\sf E}^{ b}_n}\exp\left[K_{xy}\left( \sigma_x
\xi_y + \tilde{\sigma}_x \eta_y \right)\right] . \nonumber
\end{eqnarray}
For simplicity, we do not indicate the dependence of the latter functions on the spins.
One observes that each $\Gamma_{xy} \geq 0$ due to our choice of the spin-spin interactions in (\ref{5}), cf. (\ref{Aq}),
and that $\Psi_n(\xi, \eta)$ depends only on $\sigma_x$ and $\tilde{\sigma}_x$ with $x\in {\sf S}_n$.
Fix some ${\sf E}'$ in the last line
in (\ref{M1}) and consider the subgraph ${\sf G}'$ with the edge set ${\sf E}'$ and the vertex set ${\sf V}_n$.
If in ${\sf G}'$ the vertex $z$ is disconnected from each $x\in {\sf S}_n$, then the sums over the spins
$\sigma_z, \tilde{\sigma}_z$ and over $\sigma_x, \tilde{\sigma}_x$ with $x\in {\sf S}_n$
 get independent and hence the left-hand side of (\ref{M1}) vanishes as the term $\left(\sigma_z - \tilde{\sigma}_z \right)$ is antisymmetric
 with respect to the interchange $\sigma \leftrightarrow \tilde{\sigma}$, whereas all $\Gamma_{xy}$ are symmetric and the only break of
 this symmetry is related with the fixed boundary spins
$\xi(y)$ and $\eta(y)$, $y\in {\sf S}_{n+1}$. Therefore, each non-vanishing term in
(\ref{M1}) corresponds to a path $\vartheta(z,x)$ connecting $z$ to some $x\in {\sf S}_n$.
Let us take this into account and   rewrite (\ref{M1}) as  the sum over the subsets of ${\sf E}_n$ containing the edges of at least one such path. Let
$\varTheta_n (z)$ be the set of all paths connecting $z$ to ${\sf S}_n$, and for $\vartheta \in \varTheta_n (z)$, let $\mathcal{E}_\vartheta$ be the family of subsets of ${\sf E}_n$ each of which contains ${\sf E}_\vartheta$. That is, $\mathcal{E}_\vartheta = \{ {\sf E}' \subset {\sf E}_n : {\sf E}_\vartheta \subset {\sf E}'\}$. Note that, for distinct $\vartheta, \vartheta' \in \varTheta_n (z)$, the corresponding families $\mathcal{E}_\vartheta$ and $\mathcal{E}_{\vartheta'}$ are not disjoint -- they include those ${\sf E}'$ which contain both ${\sf E}_\vartheta$ and ${\sf E}_{\vartheta'}$. Finally, set
\begin{equation}
  \label{Ma}
\mathcal{E} = \bigcup_{\vartheta \in \varTheta_n (z)} \mathcal{E}_\vartheta,
\end{equation}
that is, $\mathcal{E}$ contains all sets of edges ${\sf E}'\subset {\sf E}_n$ such that the corresponding graph ${\sf G}'$ contains at least
one path connecting $z$ to some $x\in {\sf S}_n$.
Then (\ref{M1})  takes  the form
\begin{equation}
  \label{M2a}
 M_{n,z} (\xi) - M_{n ,z} (\eta) =  \frac{1} {Z_n (\xi)Z_n (\eta)}
 \sum_{{\sf E}' \in \mathcal{E}} \sum_{\sigma_n, \tilde{\sigma}_n} \left(\sigma_z - \tilde{\sigma}_z \right)
 \Gamma ({\sf E}') \Psi_n(\xi, \eta).\end{equation}
In view of the positivity of all $\Gamma_{xy} \geq 0$, this yields
\begin{eqnarray}
  \label{M2b}
& &   \left\vert M_{n,z} (\xi) - M_{n ,z} (\eta) \right\vert  \leq  \frac{2}{Z_n (\xi)Z_n (\eta)}
 \sum_{{\sf E}' \in \mathcal{E}} \sum_{\sigma_n, \tilde{\sigma}_n}
 \Gamma ({\sf E}') \Psi_n(\xi, \eta) \\[.2cm] & & \quad
 \leq  \frac{2}{Z_n (\xi)Z_n (\eta)}\sum_{\sigma_n, \tilde{\sigma}_n}
\sum_{\vartheta \in \varTheta_n (z)} \Gamma({\sf E}_\vartheta) \sum_{{\sf E}' \in \mathcal{E}_\vartheta}\Gamma ({\sf E}'\setminus {\sf E}_\vartheta) \Psi_n(\xi, \eta).\nonumber
\end{eqnarray}
Note that the sum $\sum_{\vartheta \in \varTheta_n (z)} \sum_{{\sf E}' \in \mathcal{E}_\vartheta}$ contains the same summands as
$\sum_{{\sf E}' \in \mathcal{E}}$, see (\ref{Ma}), but a part of them are repeated, which yields the second $\leq $ in (\ref{M2b}).
From (\ref{M2}) and (\ref{Qq}) we see that $0 \leq \Gamma_{xy} \leq q(K)$, for any $x,y\in {\sf V}$.
Hence,
\[
\Gamma ({\sf E}_\vartheta) \leq [q(K)]^{\|\vartheta\|},
\]
which yields in (\ref{M2b})
\begin{eqnarray}
  \label{Mu1}
\left\vert M_{n,z} (\xi) - M_{n ,z} (\eta) \right\vert & \leq & \frac{2}{Z_n (\xi)Z_n (\eta)} \sum_{\vartheta \in \varTheta_n (z)} [q(K)]^{\|\vartheta\|} \\[.2cm] & \times & \sum_{\sigma_n, \tilde{\sigma}_n} \prod_{\{x,y\}\in {\sf E}_\vartheta}
 (1+ \Gamma_{xy})\sum_{{\sf E}''\subset {\sf E}_n \setminus {\sf E}_\vartheta} \Gamma ({\sf E}'') \Psi_n(\xi, \eta) \nonumber \\[.2cm]
 & = & 2  \sum_{\vartheta \in \varTheta_n (z)} [q(K)]^{\|\vartheta\|}. \nonumber
\end{eqnarray}
Here we have taken into account that, see (\ref{M1}),
\begin{eqnarray*}
& &   \sum_{\sigma_n, \tilde{\sigma}_n} \prod_{\{x,y\}\in {\sf E}_\vartheta}
 (1+ \Gamma_{xy})\sum_{{\sf E}''\subset {\sf E}_n \setminus {\sf E}_\vartheta} \Gamma ({\sf E}'') \Psi_n(\xi, \eta)\\[.2cm]
& & \quad  =  \sum_{\sigma_n, \tilde{\sigma}_n} \prod_{\{x,y\}\in {\sf E}_n} (1+ \Gamma_{xy}) \Psi_n(\xi, \eta) =  Z_n (\xi)Z_n (\eta).
\end{eqnarray*}
Note that a similar construction was used in deriving  (2.16) and (2.17) in \cite{Dr}, see also Lemma 3.1 in \cite{KK}.
Since $n_z$ is defined by the condition $z\in {\sf S}_{n_z}$,  the shortest path in $\varTheta_n (z)$ has length
$n-n_z$. Furthermore, for every $x\in {\sf S}_n$, there exists exactly one path $\vartheta \in \varTheta_n(z)$. Then, for $q(K) \leq 1$,
the estimate (\ref{R2}) follows from (\ref{Mu1}).

\subsection{The proof of Lemma \ref{grflm}}

If we prove that, for $c>s^\gamma$, with probability one
\begin{equation}
  \label{Lh4}
\sum_{n\geq 1} c^{-n} Y_n < \infty,
\end{equation}
see (\ref{Lh1}), then the proof that $L_n/c^n\to W \in [0,+\infty)$,  with probability one, follows as in the proof of Theorem 6.1 in \cite[page 50]{As}. The first summand of $L_n$ in (\ref{Lh}) is under control in view of the assumed property
$c\geq a-1$.

As $\alpha\in (0,1)$, by the standard Minkowski inequality, cf. \cite[Theorem 8, page 319]{GS}, as well as by (\ref{Lh3}) and (\ref{Log}), we have
\[
 \langle Y_n^\alpha \rangle \leq \sum_{j=1}^{s^{n-1}} \bigg{
 \langle} \left(\widehat{X}_j - s \right)_+^\alpha \bigg{\rangle} = s^{n-1} b_\alpha.
\]
Then, for any $q>c^\gamma$, by Markov's inequality, cf. \cite[page 311]{GS},
\begin{equation}
  \label{Lh5}
P_n:= {\rm Prob}\left( q^{-n} Y_n > 1\right) = {\rm Prob}\left( Y_n^\alpha > q^{\alpha n}\right)   \leq \frac{b_\alpha}{s} \left( \frac{s}{q^\alpha}\right)^n.
\end{equation}
On the other hand, for $c> s^\gamma$, we take $q\in (s^\gamma, c)$ and rewrite
\begin{equation}
  \label{Lh6}
\sum_{n\geq 1} c^{-n} Y_n = \sum_{n\geq 1} \left(\frac{q}{c}\right)^n \left( q^{-n} Y_n \right).
\end{equation}
For such $q$, by (\ref{Lh5}) we have  that
$\sum_{n\geq 1} P_n < \infty$. Thus, the Borel-Cantelli lemma, cf. \cite[page 320]{GS}, yields that with probability one
only finitely many of the events  $q^{-n} Y_n > 1$
may occur, which by (\ref{Lh6}) readily yields (\ref{Lh4}).

\section{Concluding comments and remarks}
\label{5S}

\subsection{The graph model}

Let us look at the structure of generation $n$ in  ${\bf GW}(s,p)$, cf. (\ref{Lh}) -- (\ref{Lh3}).
For $s=1$, we deal with a size-biased Galton-Watson tree, cf. \cite{Yuval} and \cite{LyP}, and
 $b< \infty$ is  known as the $X \log X$ condition of the Kesten-Stigum theorem, see also
\cite[page 23]{As}. If it holds, the influence of the distinguished ray on the structure of ${\sf S}_n$ is asymptotically negligible,
see Section 3 in \cite{LyP}.
For $s\geq 2$, the number of distinguished vertices $\widehat{L}_n$ in generation $n$ increases in such a way that
the sequence $\{\sigma^{-n}\widehat{L}_n\}_{n\in \mathbb{N}}$ with probability one tends to a random variable $\widehat{W}$ such that
${\rm Prob} (\widehat{W} > 0) =1$ and $\langle \widehat{W} \rangle =1$. Here
\[
\sigma = s - (s-1)\hat{p}_2 - (s-2)\hat{p}_3 - \cdots - \hat{p}_s.
\]
Note that $\sigma \in (1, s]$, and $\sigma =s$ if and only if $p_2 = \cdots = p_s =0$, and hence the least value of $X$ is $s$.
In this case, $a-1>s$.
If $a-1> s^\gamma$, $L_n$ asymptotically `behaves' like in the ordinary Galton-Watson tree, which means that the production of offsprings
with probability $p$ `dominates' the immigration  from the distinguished part of the population, which is always the case for ${\bf GW}(1,p)$, cf. (\ref{Aa}). However, for
$a-1\leq s^\gamma$, the main contribution into $L_n$ comes from the immigration. If $a-1$ is close to one and
$s^\gamma\gg 1$, even for big $n$ the  structure of $({\sf V}_n, {\sf E}_n)$ is quite close to that of the corresponding subgraph
of ${\bf GW}(0,\hat{p})$, i.e., as in the case of the configuration model. Note that $s^\gamma$ is  big whenever (a) $s$ is big, and hence
the distinguished subgraph is `big'; (b) $\alpha$ is small, and hence the offspring production in
the distinguished subgraph is very intensive.

Noteworthy, the procedure of formation of edges in ${\bf GW}(s,p)$ resembles that of \cite{BCK,BK}. The `value' specified in \cite{BK} by
a positive $\omega$, in our case appears as a mark $k$, distributed according to $p$ with $p_k$ `decreasing at infinity', cf. ``inverse mass-action principle" of \cite{BK}.
Distinguished individuals have the right to choose neighbors.  Each chooses according to its own mark at most $s$ distinguished neighbors,
preferring those with big marks. The restriction of the number of distinguished neighbors to $s$  corresponds to the restriction of the number of outgoing edges in the Cameo principle \cite{BK}.

One observes that ${\bf GW}(s,p)$ is a natural generalization of the size-biased Galton-Watson tree ${\bf GW}(1,p)$.
In a separate work, we plan to study this model in more detail. Another generalization of ${\bf GW}(1,p)$ can be the model in which
the distinguished part of $\widehat{X}$ is not $\widehat{X}- (\widehat{X}-s)_{+}$ like in
${\bf GW}(s,p)$, cf. (\ref{Lh}), but some more general increasing function of $\widehat{X}$, or is random. Further generalizations of this kind can be obtained in the approach of \cite{Jag}.

\subsection{The thermodynamic states}
\label{conc}

In 1989, R. Lyons \cite{Rly} proved that, for any tree, the critical temperature of the model (\ref{1}) with $h=0$ and $J_{xy} = J >0$ is
\begin{equation}
  \label{rl}
T_c = \frac{J}{k_B \coth^{-1} \varrho},
\end{equation}
where $\varrho$ is the {\it branching number} of the underlying tree. For an infinite (almost surely non-extinct) Galton-Watson tree, with probability one $\varrho$ equals
the mean number of offsprings, see Proposition 6.4 in \cite{LA}. Hence,
for ${\bf GW}(0,\hat{p})$, we have $\varrho= \langle k^2 \rangle/\langle k \rangle -1$, which yields in (\ref{rl})
\begin{equation}
  \label{3}
T_c = { 2J} \bigg{/}{k_B \ln  \frac{ \langle k^{2} \rangle }{ \langle k^{2} \rangle - 2\langle k \rangle}}.
\end{equation}
The very same expression for $T_c$ was `rediscovered' in 2002 in \cite{D,L}. It is significant that Lyon's paper \cite{Rly} was
quoted\footnote{In the preprint arXiv:0705.0010 version  of \cite{Dor}, both papers \cite{Rly,LA} were quoted.}
in \cite{Dor} and even the same formula (\ref{rl})  was discussed, see the text between equations (89) and (90) on page 1304 in \cite{Dor}
or the text between equations (89) and (90) on page 33 in arXiv:0705.0010.

Now let us turn to the discussion of  Theorem \ref{Tm}, which
describes the model (\ref{1}) with {\it arbitrary} signs\footnote{The antiferromagnetic case $J_{xy}= -J$, $J>0$, can be reduced to
the ferromagnetic one by changing signs of $\sigma_x$, $x\in {\sf S}_{2n}$.}  of the intensities $J_{xy}$. The only condition is that
they are uniformly bounded, cf. (\ref{Aq}). As follows from the estimate (\ref{R2}), the model (\ref{1}) on a general tree
can be in a paramagnetic state if the number of vertices in the sphere ${\sf S}_n$ admits the control $|{\sf S}_n|\leq c^n$ for some $c>1$.
Note that in the example of \cite{KK}, $|{\sf S}_n| \sim n!$.
For graphs other than trees\footnote{For more on phase transitions on graphs see \cite{LM}.}, one has to control $\varTheta_n (z)$, cf. (\ref{Mu1}).
The results of Theorem \ref{Tm} can naturally be extended to models with continuous spins and bounded interaction,
as e.g. in \cite{KK}.
In a separate work, we plan to study the case of unbounded interaction intensities, which includes also random $J_{xy}$, as e.g. in the Edwards-Anderson model. For a ferromagnetic Ising model on a ${\bf GW}(0,p)$, the exact value of the critical
temperature is given in (\ref{rl}) with $\varrho = a-1$.
The corresponding result of Theorem \ref{Tm} is that the  model is in a paramagnetic state for
\[
T >  { 4J}\bigg{/}{k_B \ln  \frac{ a }{ a - 1}} > T_c.
\]
It is naturally less precise as we cover the case of arbitrary signs of $J_{xy}$.

Regarding the model ${\bf GW}(0,\hat{p})$ studied in \cite{D,L}, by Theorem \ref{1} we get that the Ising model with  arbitrary signs
of $J_{xy}$ obeying (\ref{Aq}) is in a paramagnetic state if
\[
T >  { 4J}\bigg{/}{k_B \ln  \frac{ \langle k^{2} \rangle }{ \langle k^{2} \rangle - 1}}.
\]
It is believed, cf. \cite{Cohen}, that the Internet can well be modeled as a scale-free graph with $\lambda =5/2$, which
corresponds to the choice of $\alpha < 1/2$, see (\ref{Log}). As a byproduct of the result of Theorem \ref{Tm},
we obtain the following, cf. \cite{Cohen,Cohen1},
\begin{proposition}
  \label{Cpn}
Let $\theta\in (0,1)$ be the Bernoulli bond percolation probability on ${\bf GW}(s,p)$ graphs with $s\geq 2$ and $p$ obeying (\ref{PQ}) and $b_\alpha < \infty$.
Then with probability one there is no giant component if
\[
\theta < \min\{q(K_c); q(\widehat{K}_c)\},
\]
where $q(K)$ is given by (\ref{Qq}), and $K_c$ and $\widehat{K}_c$ are as in Theorem \ref{1}.
\end{proposition}

\section{Acknowledgment}

The author benefited from the discussions on the matter of this work
with Philippe Blanchard, Yurij Holovatch, and Yuri Kondratiev, for that he is
cordially indebted. The work was supported in part by the DFG through SFB 701: ``Spektrale Strukturen
und Topologische Methoden in der Mathematik"  and through the
research project 436 POL 125/113/0-1.



\end{document}